\pdfoutput=1

\documentclass[11pt,twoside,a4paper,cmspaper,final,collab]{cms-tdr}
\def\svnVersion{412499}\def\cmsCernNoTag{CERN-EP/2017-029}\def\cmsCernDate{\today}\def\cmsMessage{Published in Physics Letters B as \href{http://dx.doi.org/10.1016/j.physletb.2017.06.009}{\doi{10.1016/j.physletb.2017.06.009.}}}

\begin{document}\cmsNoteHeader{SMP-13-008}

\hyphenation{had-ron-i-za-tion}
\hyphenation{cal-or-i-me-ter}
\hyphenation{de-vices}
\RCS$Revision: 305398 $
\RCS$HeadURL: svn+ssh://dudero@svn.cern.ch/reps/tdr2/notes/SMP-13-008/trunk/SMP-13-008.tex $
\RCS$Id: SMP-13-008.tex 305398 2015-10-06 23:50:48Z dudero $
\newlength\cmsFigWidth
\ifthenelse{\boolean{cms@external}}{\setlength\cmsFigWidth{0.85\columnwidth}}{\setlength\cmsFigWidth{0.4\textwidth}}
\ifthenelse{\boolean{cms@external}}{\providecommand{\cmsLeft}{top}}{\providecommand{\cmsLeft}{left}}
\ifthenelse{\boolean{cms@external}}{\providecommand{\cmsRight}{bottom}}{\providecommand{\cmsRight}{right}}
\newcommand{\mt}{\ensuremath{m_{\mathrm{T}}}\xspace}
\newcommand{\ptJ}{\ensuremath{\pt^J}\xspace}
\newlength{\columnfigure}
\newlength{\tripleColFig}
\ifthenelse{\not{\boolean{cms@external}}}{%
  \setlength{\columnfigure}{0.65\columnwidth}
  \setlength{\tripleColFig}{0.32\columnwidth}
}
{%
  \setlength{\columnfigure}{0.9\columnwidth}
  \setlength{\tripleColFig}{0.68\columnwidth}
}
\cmsNoteHeader{SMP-13-008}
\title{Search for anomalous couplings in boosted $\PW\PW/\PW\Z\to{}\ell\nu\qqbar$ production in proton-proton collisions at $\sqrt{s}=8\TeV$}

\date{\today}

\abstract{
This Letter presents a search for new physics manifested as
anomalous triple gauge boson couplings in $\PW\PW$ and $\PW\cPZ$ diboson
production in proton-proton collisions. The search is performed using
events containing a \PW\ boson that decays leptonically
and a \PW\ or \cPZ\ boson whose decay products are merged into a single reconstructed jet. The data,
collected at $\sqrt{s}=8\TeV$ with the CMS detector at the
LHC, correspond to an integrated luminosity of 19\fbinv.  No
evidence for anomalous triple gauge couplings is found and the
following 95\% confidence level limits are set on their values:
$\lambda$\,($[-0.011,0.011]$),
$\Delta{\kappa_\gamma}$\,($[-0.044,0.063]$), and
$\Delta{g_1^{\Z}}$\,($[-0.0087,0.024]$).  These limits are also translated
into their effective field theory equivalents:
$c_{\PW\PW\PW}/\Lambda^2$\,($[-2.7,2.7] \TeV^{-2}$),
$c_\mathrm{B}/\Lambda^2$\,($[-14,17] \TeV^{-2}$), and
$c_{\PW}/\Lambda^2$\,($[-2.0,5.7] \TeV^{-2}$).
}

\hypersetup{%
pdfauthor={CMS Collaboration},%
pdftitle={Search for anomalous couplings in semileptonic WW/WZ to l nu qqbar  production  in proton-proton collisions at sqrt(s)=8 TeV},%
pdfsubject={CMS},%
pdfkeywords={CMS, physics, aTGC}}

\maketitle

\section{Introduction}
  Measurements of electroweak diboson
production can be translated into measurements of gauge boson
self-couplings,
which are among the most fundamental aspects of the standard model (SM).  At
leading order (LO), only $s$-channel $\cPq\cPaq$ annihilation diagrams have
a triple-boson vertex. In \PW\PW\ production, the $\PW\PW\gamma$ and $\PW\PW\cPZ$ vertices
contribute, while in $\PW\cPZ$ production only the \PW\PW\cPZ\ vertex is present.
Physics beyond the SM can modify the couplings at these vertices, leading to
observable differences in the cross section and the kinematic
distributions of final state particles~\cite{Dixon:1999di}.
In the search for anomalous triple gauge couplings (aTGCs), we adopt the effective
Lagrangian and LEP param\-etrization in Ref.~\cite{PhysRevD.48.2182}, without form factors:
$\lambda_{\gamma} = \lambda_{\cPZ} = \lambda$,
$\Delta{\kappa_{\cPZ}} = \Delta{g_1^{\cPZ}}-\Delta{\kappa_\gamma} \, \tan^2\theta_{{\PW}}$.
We focus in particular on the parameters $\lambda$,
$\Delta{\kappa_\gamma}$, and $\Delta{g_1^{\cPZ}}$, where the deltas represent deviations from
their respective SM values ($\lambda_\mathrm{SM}=0$).
We also translate these
into the equivalent parameters defined in an effective field
theory (EFT) approach, namely $c_{\PW\PW\PW}/\Lambda^2$, $c_{\PW}/\Lambda^2$, and
$c_\mathrm{B}/\Lambda^2$, where $\Lambda$ is the scale of new physics~\cite{Degrande:2012wf}.

This Letter presents a search for new physics
manifested as anomalous couplings of triple gauge boson vertices in \PW\PW\ or \PW\cPZ\ diboson
production from pp collisions at $\sqrt{s}=8\TeV$ at the CERN LHC.
We focus on the case where one \PW\ boson
decays leptonically (\PW$_\text{lep}\to\ell\nu$, with $\ell=\Pe,\Pgm$),
while the other vector boson V$_\text{had}$ decays hadronically, giving
rise to a single merged jet ($J$) in the final state.  Previous searches in
this channel at the LHC can be found in
Refs.~\cite{Chatrchyan:2012bd,Aad:2014mda}. Other recent searches in the leptonic
channel are described in Refs.~\cite{Khachatryan:2015sga,Aad:2016wpd}.  The advantages
of reconstructing \PW{}V pairs in the $\ell\nu \qqbar$ decay mode over
purely leptonic final states are the larger branching fractions
of \PW\ and \cPZ\ bosons to quarks, and in the case of two \PW\ bosons, the ability to
reconstruct their transverse momenta (\pt). These advantages are partially offset
by the larger backgrounds in the $\ell\nu \qqbar$ channel, arising mainly
from \PW+jets production. The sensitivity
of \PW\PW\ production to the \PW\PW$\gamma$ coupling and of
both \PW\PW\ and \PW\cPZ\ production to the \PW\PW\cPZ\ coupling,
especially at high boson \pt, makes these processes
particularly useful as a probe of aTGCs.

Compared to our previous search at $\sqrt{s}=7\TeV$~\cite{Chatrchyan:2012bd},
we have added another coupling parameter, $\Delta{g_1^{\cPZ}}$, to the parameter space,
and we focus exclusively on the Lorentz-boosted final states, where V$_\text{had}$ is
reconstructed as a single merged jet, since these final states are far
more sensitive to an aTGC signal than the resolved two-jet states.

\section{The CMS detector}

The central feature of the CMS apparatus is a superconducting solenoid
of 6\unit{m} internal diameter, providing a magnetic field of
3.8\unit{T}. Within the solenoid volume are a silicon
pixel and strip tracker, a lead tungstate crystal electromagnetic
calorimeter (ECAL), and a brass and scintillator hadron calorimeter,
each composed of a barrel and two endcap sections. Muons are measured
in gas-ionization detectors embedded in the steel flux-return yoke
outside the solenoid. The CMS detector is nearly hermetic, allowing
for measurements of the missing transverse momentum (\MET) in the
event. \MET is defined as the magnitude of the negative vector \pt sum of
all reconstructed particles in an event. A
two-tier trigger system selects the events of interest.  A more
detailed description of the CMS detector, together with a definition
of the coordinate system used and the relevant kinematic variables,
can be found in Ref.~\cite{Chatrchyan:2008zzk}.

\section{Data and simulation samples}

The data were collected using single-lepton triggers with
\pt thresholds of 24\,(27)\GeV for muons (electrons).  The
overall trigger efficiency is about 94\% (90\%) for the muon
(electron) data, with a small dependence (a few percent) on \pt and
pseudorapidity $\eta$.  The total integrated luminosity collected and
processed is 19.3 (19.2)\fbinv for muon (electron) triggers.

We use the
{\MADGRAPH}5~~1.3.30~\cite{madgraph5} event generator to produce both
the \PW+jets and Drell--Yan samples, with up to four additional partons in the matrix element calculation.  Single top quark and top
quark-antiquark pair (\ttbar) samples are generated with
\POWHEG~1.0~\cite{Nason:2004rx,Frixione:2007vw,Alioli:2010xd,Frixione:2007nw,Alioli:2009je}.
The diboson samples (\PW\PW, \PW\cPZ) are
generated on-shell at next-to-LO (NLO) with {\MADGRAPH}5\_a\MCATNLO version 2.0.0~\cite{Alwall:2014hca}
and \textsc{MadSpin} version 3.2~\cite{Artoisenet:2012st}. The
decays $\PW\to\Pgt\nu$ are included for
all processes.  The \Pgt\ lepton decays are simulated with \TAUOLA~\cite{Jadach:1993hs}.
The \PYTHIA 6.422 generator~\cite{PYTHIA} provides the fragmentation and parton
shower simulation, with the parameters of the underlying
event set to the Z2* tune~\cite{TuneZ2star1,TuneZ2star2}.
The \kt-MLM matching scheme is used to interface {\PYTHIA}6 with {\MADGRAPH}5 at LO~\cite{Alwall:2007fs}.
The set of parton
distribution functions (PDFs) used is CTEQ6L1~\cite{Pumplin:2002vw}
for LO generators and CT10~\cite{Lai:2010vv} for NLO generators.
A \GEANTfour-based simulation~\cite{GEANT4} of the CMS detector is
used in the production of all Monte Carlo (MC) samples.
The simulation also includes multiple proton-proton collisions within a bunch crossing (pileup).
Simulated events are reconstructed and analyzed in the
same way as measured collision events, subject to additional
corrections that account for differences between data and simulation
in trigger and selection efficiencies, and in the
vertex multiplicity distribution.

\section{Event reconstruction}

All observable objects, namely leptons, jets, and \MET, are
reconstructed with a particle-flow
technique~\cite{CMS-PAS-PFT-09-001,CMS-PAS-PFT-10-001} that combines
information from several subdetectors. Muons are reconstructed within
$\abs{\eta}<2.4$ with the inner tracker and the muon
system~\cite{Chatrchyan:2013sba}.  Electrons are reconstructed within
$\abs{\eta}<2.5$ from
tracks in the tracker pointing to energy clusters in the ECAL, and identified using a
multivariate discriminator~\cite{Khachatryan:2015hwa}. The selections applied to this
discriminator are tuned to match the $\eta$-binned efficiencies used for
Ref.~\cite{Chatrchyan:2012bd}.  Muons (electrons) are required to have \pt
greater than 25\,(30)\GeV. The lepton candidates are required to be
consistent with originating from the event's primary vertex,
and to be isolated from other activity in the event. The
isolation requirements for muons (electrons) are based on the
particle-flow technique with an isolation cone of $\Delta R
= 0.4$\,(0.3), and are designed to reduce the effects of pileup
and neutral particles. Events with additional loosely identified leptons are
vetoed to reduce the backgrounds from fully leptonic decays, such as
those originating from the Drell--Yan process and diboson production.
Decays of the tau lepton to electrons or muons that pass these
criteria are included as potential signal events.

The anti-\kt (AK)~\cite{antikt,Cacciari:2011ma} and
Cambridge--Aachen (CA)~\cite{antikt,Cacciari:2011ma,Dokshitzer:1997in}
clustering algorithms are used to reconstruct jets in the event. The AK
algorithm uses a distance
parameter of $R=0.5$ (AK5). The CA jets are
clustered with $R=0.8$ (CA8) and are used for
reconstructing V$_\text{had}$, where the V boson decay products are
merged into a single jet. The combined secondary vertex algorithm at the medium
operating point is used to tag AK5 jets as
b jets~\cite{Chatrchyan:2012jua}. We assign the \MET measured in the
event to the neutrino candidate and combine this with the identified lepton
to reconstruct \PW$_\text{lep}$. Boosted \PW\ events are selected by
requiring $\pt>200\GeV$ for \PW$_\text{lep}$.

We require one CA8 jet with $\pt>200\GeV$, and no additional CA8 jets
with $\pt>80\GeV$, in the region $\abs{\eta}<2.4$. The \MET is required
to be above 50 (70)\GeV for the muon (electron) channel to
suppress multijet backgrounds. We ensure that the two bosons are
back-to-back by requiring $\Delta R(\ell,J)>\pi/2$,
$\Delta\phi(\MET,J) > 2.0$, and $\Delta\phi(\PW_\text{lep},J) > 2.0$.
We veto events based on the
presence of any b-tagged AK5 jets with $\pt>20\GeV$ and outside the CA8 jet cone
to reduce the \ttbar\ background. After the kinematic
selections, we apply jet substructure techniques. Improved separation
between the signal and the multijet background is obtained in the jet mass
observable by means of a ``pruning'' algorithm~\cite{SMPJS,
Ellis:2009me} designed to remove soft gluon radiation and pileup contributions
from jets.
The ``$N$-subjettiness'' variable~\cite{Thaler:2010tr} is a jet
substructure observable that defines a measure, $\tau_N$, for a jet to
have $N$ subjets.  We require $\tau_2 / \tau_1$, which is the ratio of
2-subjettiness to 1-subjettiness, of the leading CA8 jet to be less
than 0.55 to discriminate against \PW+jets backgrounds.

\section{Background and signal modeling}

After all selections the background comprises three main components:
\PW+jets, top quark (\ttbar and single top quark), and SM diboson production.
Multijets, \cPZ+jets, \cPZ\cPZ, \cPZ$\gamma$, H(125)$\to\PW\PW^*$, and fully hadronic and leptonic
\PW\PW\ decay mode backgrounds were estimated and determined to be negligible.

For the aTGC search we select the merged jet \pt, $\ptJ$,
as the observable, which for diboson pairs is the \pt of V$_\text{had}$. We take the
binned shape of the $\ptJ$ distribution for each contributing process from
MC samples. However, since the LO \PW+jets prediction falls below
the data, we choose to extract the normalizations
of the largest background components first from an unbinned
maximum-likelihood fit to the data distribution of the merged jet mass,
$m_J$. The diboson $m_J$ shape in the fit region is
unaffected by the aTGC signal at the level of sensitivity of this analysis.

\begin{figure*}[tbh]
\includegraphics[width=\tripleColFig]{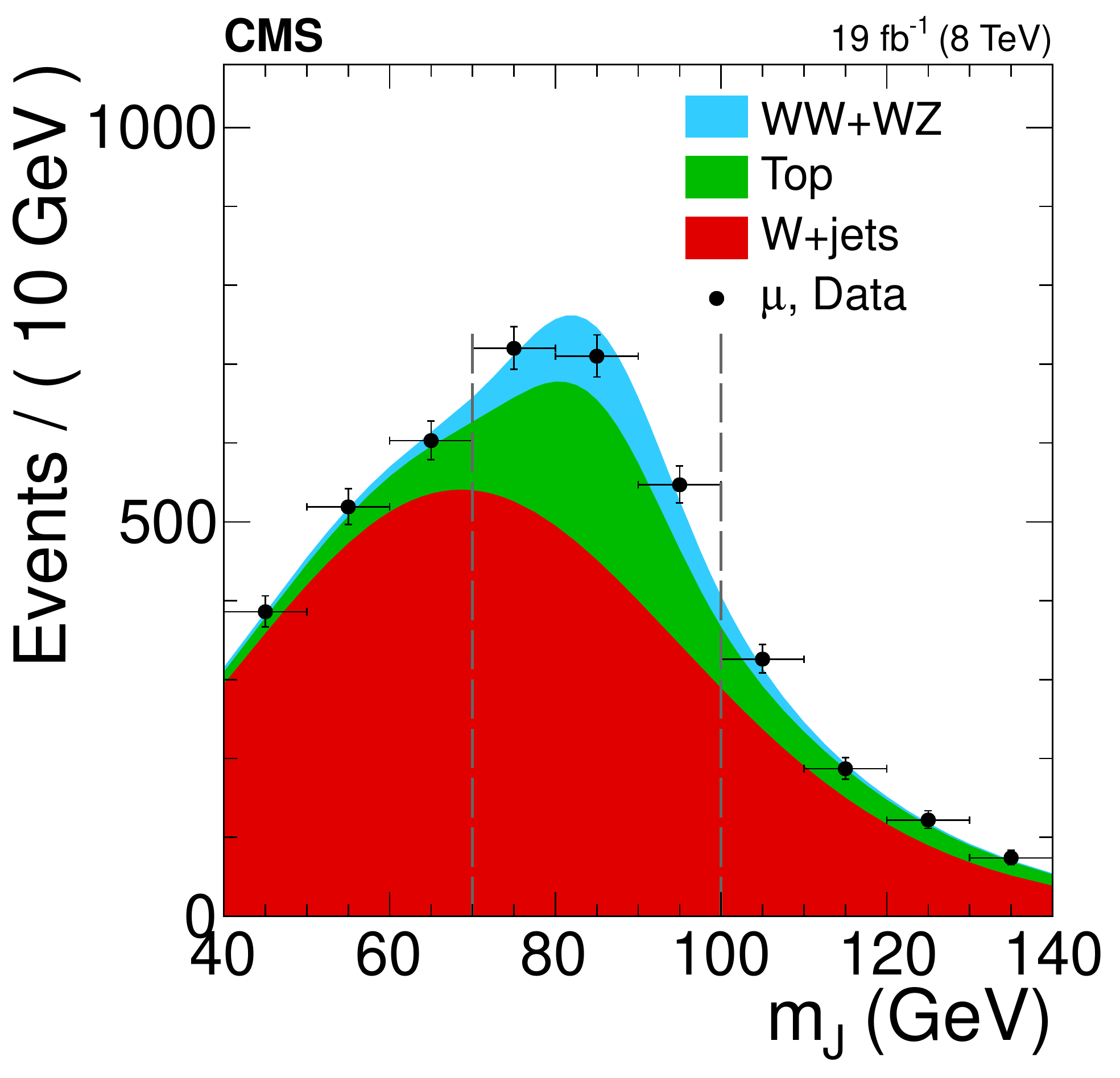}
\includegraphics[width=\tripleColFig]{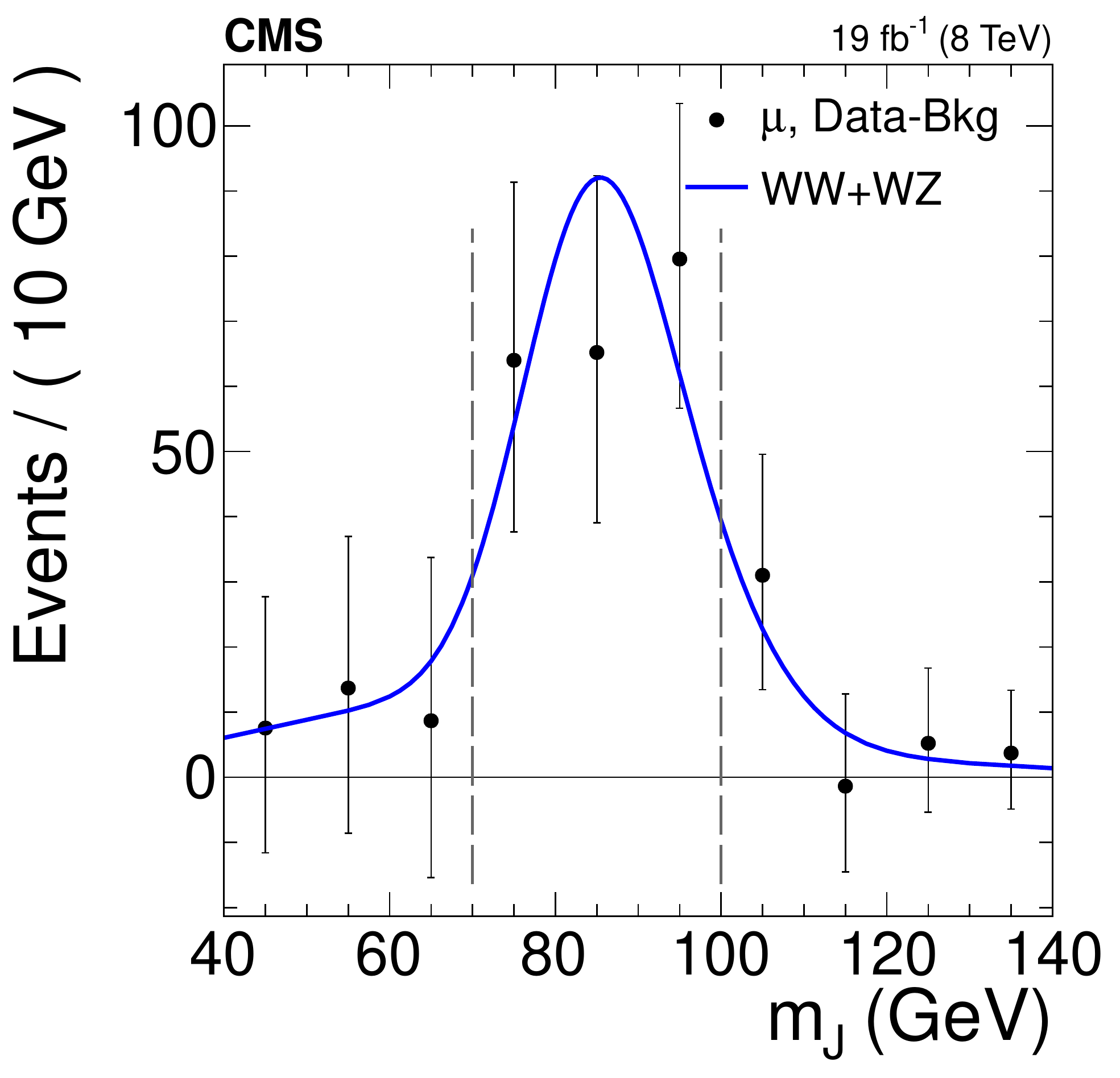}
\includegraphics[width=\tripleColFig]{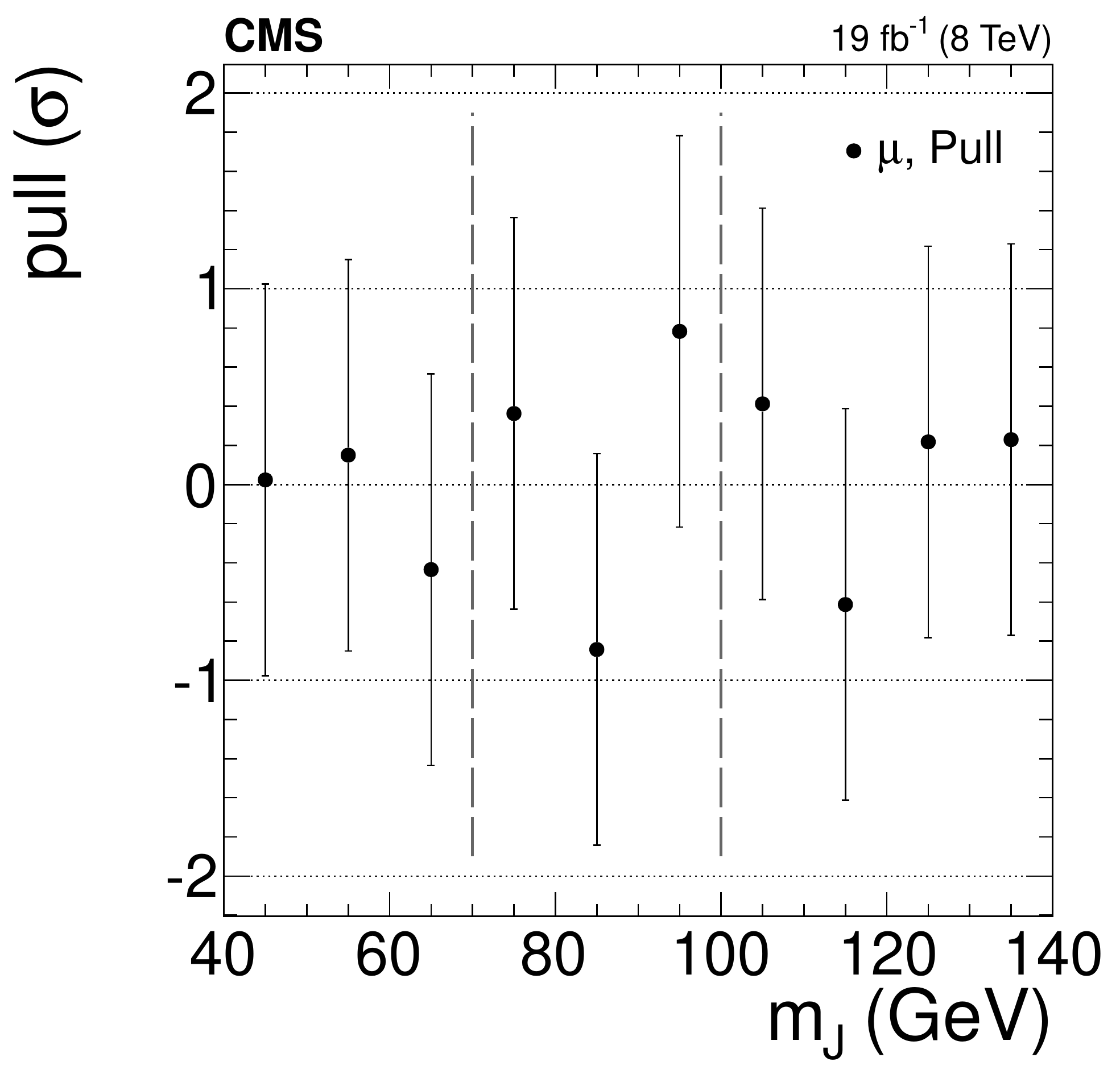}\\
\includegraphics[width=\tripleColFig]{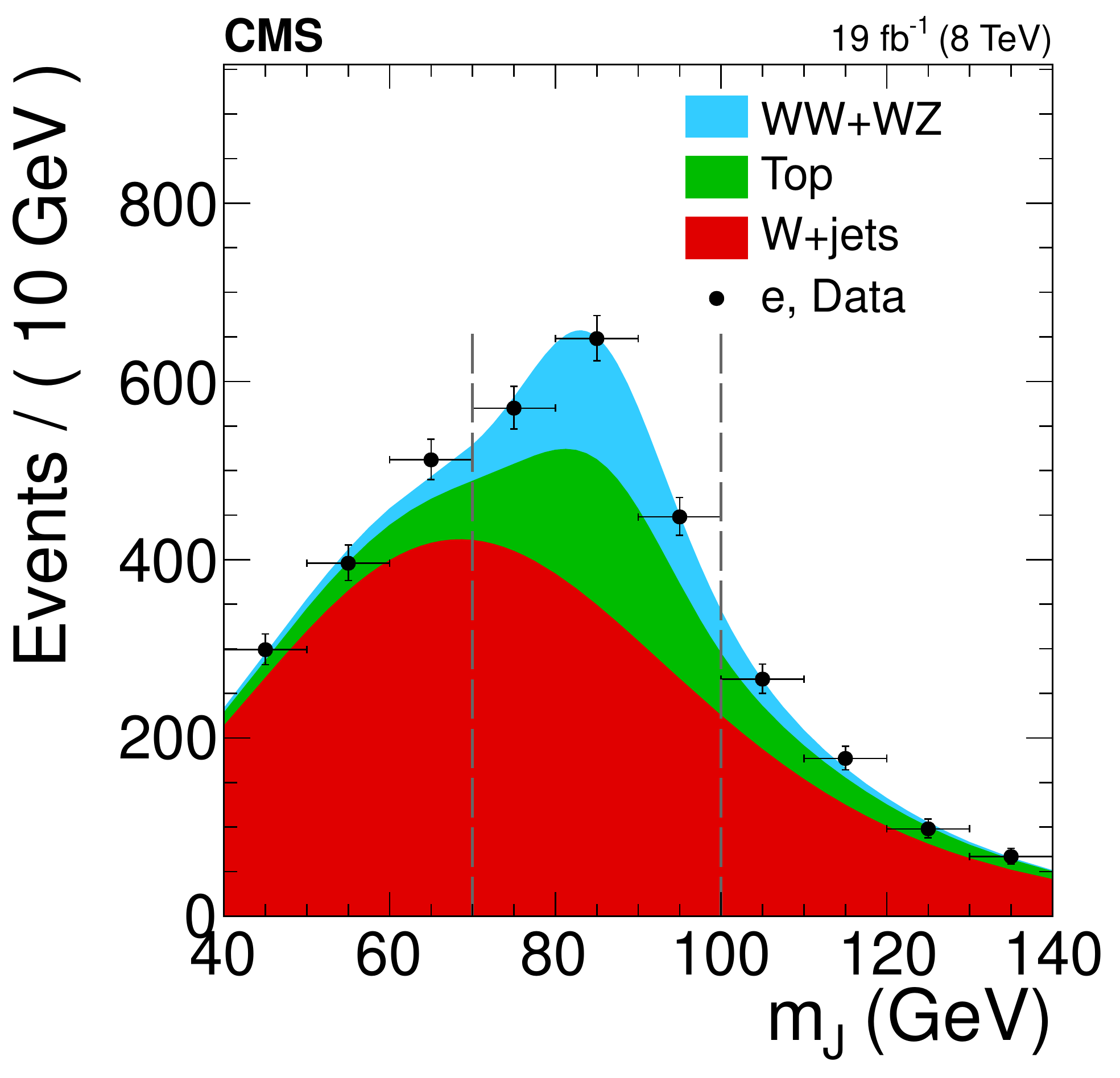}
\includegraphics[width=\tripleColFig]{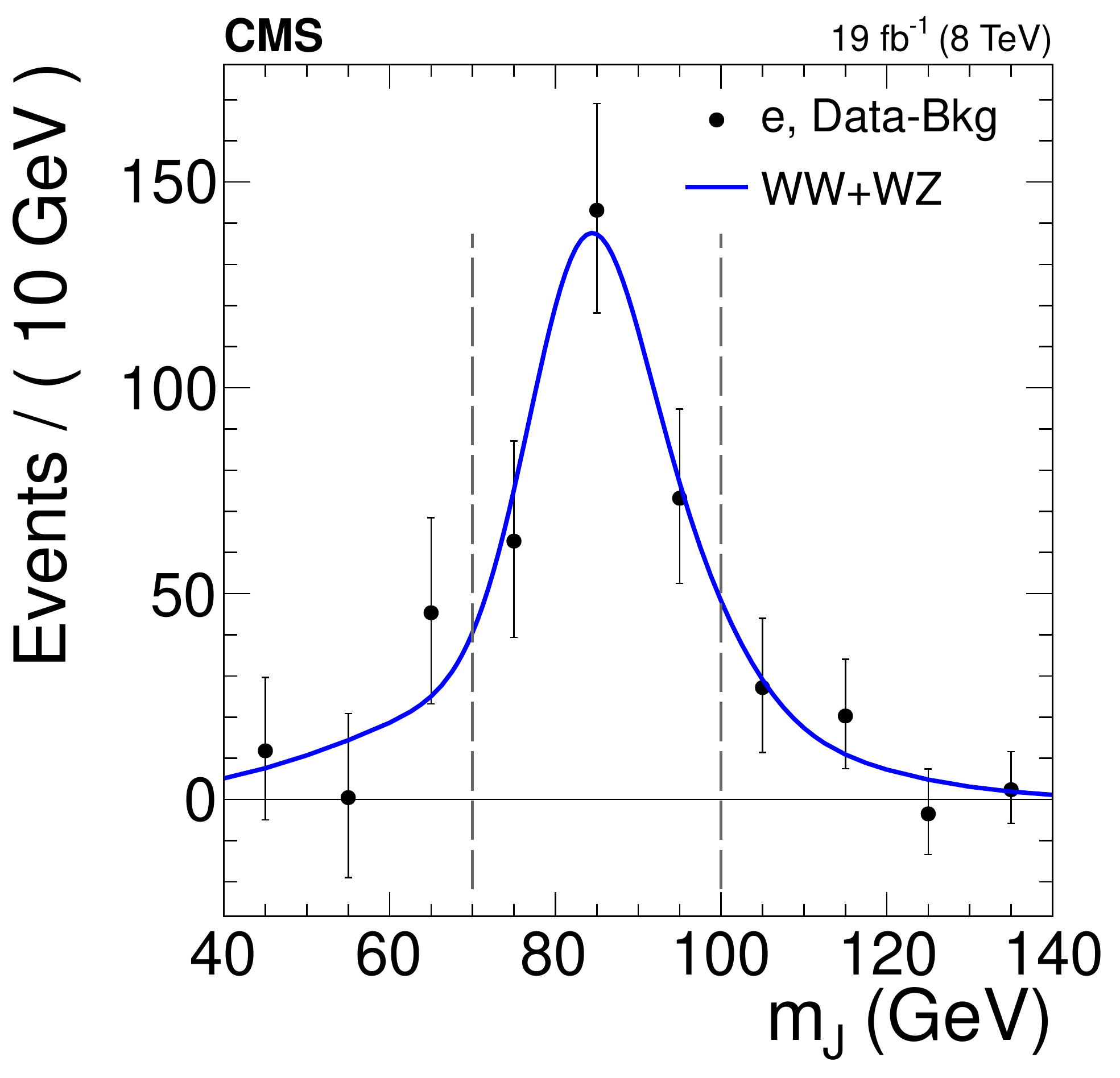}
\includegraphics[width=\tripleColFig]{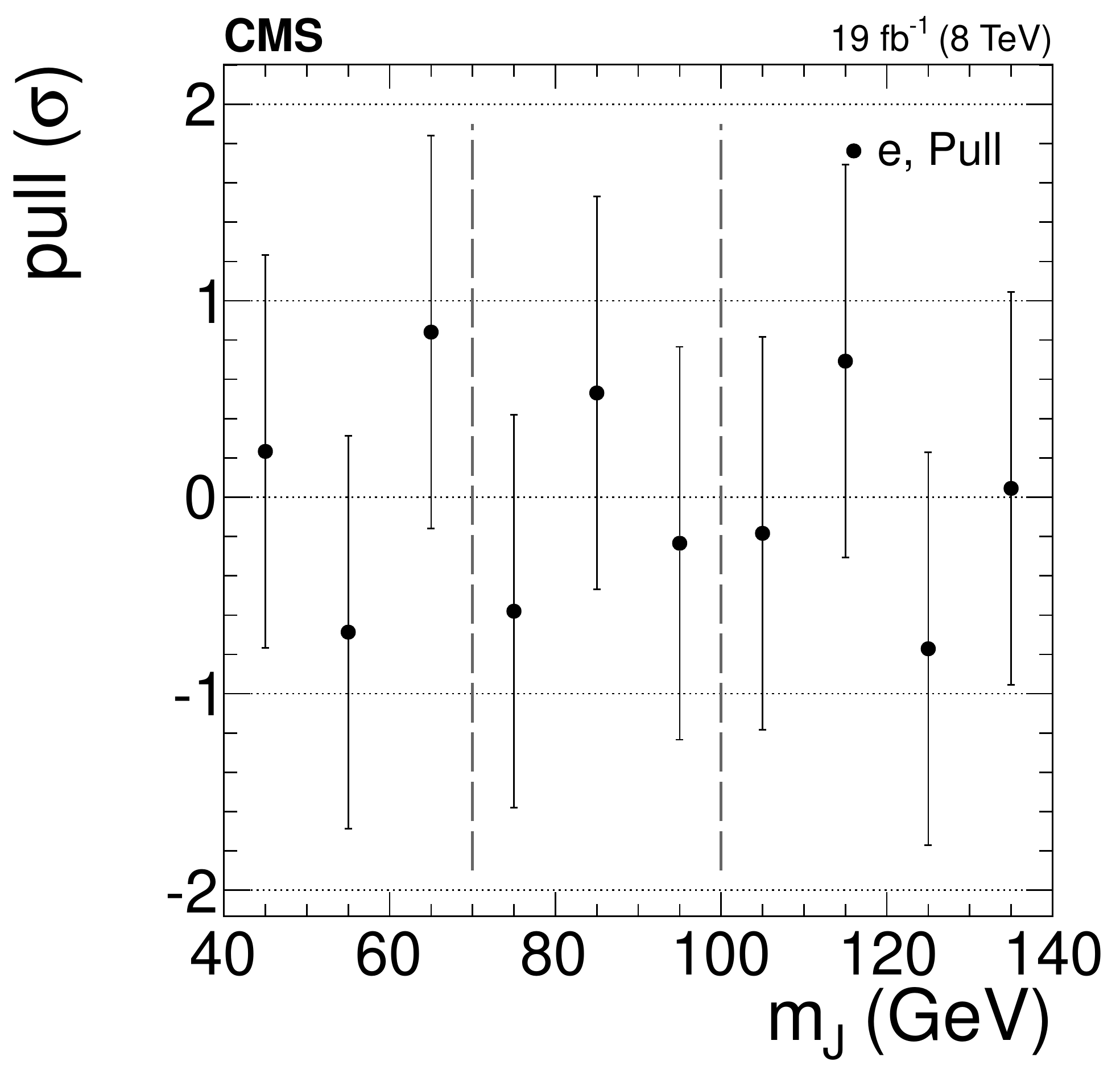}
\caption{
Post-fit distributions of the merged jet invariant mass for muons (top) and
electrons (bottom) with the estimates of the relevant backgrounds.
The merged jet invariant mass is plotted for all events
(left), after subtraction of all components except the diboson
(center), and the subsequent normalized residual or pull distributions:
$(\text{data} - \text{fit})/(\text{fit uncertainty})$ (right). The
error bars represent statistical uncertainties. The dashed vertical
lines mark the signal region of $70<m_{J}<100\GeV$, from which the
the \pt distribution normalizations are extracted. }
\label{fig:Fig1}
\end{figure*}

\subsection{Normalization extractions from the \texorpdfstring{$m_J$}{merged jet mass} fit}

For this part of the analysis we employ a two-stage procedure: first we fit the distribution in
simulation for each process individually. The MC templates used in the 7\TeV analysis are
replaced by analytical functions, which provide additional flexibility to model the data accurately.
Second, we utilize the results from the first set of
fits to perform an unbinned maximum-likelihood fit to data that includes
all components. Due to the differences in background
compositions and shapes, the fit to data is performed separately for the
muon and electron channels. All fits are performed over the mass range
$40<m_{J}<140\GeV$. Within each fit to data, the normalization for each
background process is either free to float
or allowed to vary around a central value subject to a Gaussian
constraint. Some components have been combined because of similarity in shape,
or because the \PW\ and \cPZ\ bosons are not well-resolved in $m_J$.
Finally, the
yields used to normalize background $\ptJ$ components are
extracted from the signal region of $70<m_{J}<100\GeV$.

To assist in the background determination, we define a control sample
intended to isolate pure top quark events for comparison with
simulation~\cite{Khachatryan:2014gha}.  The sample is constructed by
inverting the selection on the number of b-tagged AK5 jets outside the
CA8 jet, thus requiring that there be at least one AK5 b-tagged jet.
This control sample is subsequently referred to as the top control
sample.

The diboson probability density function (pdf) in $m_J$ is
parametrized by a sum of two Gaussian functions
corresponding to the \PW\ and \cPZ\ resonances. The position and
width of the \cPZ\ Gaussian are fixed with respect to those of the
\PW\ Gaussian, which is initially taken from simulation. The relative
fractions of \PW\PW\ (84\% of the total) and \PW\cPZ\ (16\%) are also taken
from simulation.
The broad background from jets misassigned to V$_\text{had}$ is
modeled by an error function times an exponential function.  The \PW\ Gaussian
parameters are
subsequently corrected with MC-to-data scale factors determined from the
top control sample, in order to account for mismodeling of the merged-jet
mass in simulation.  All diboson shape parameters are then fixed during the fits to
the data, while the normalizations are free parameters to be measured.

For the \PW+jets process, the shape of the $m_{J}$ distribution is
described by a kinematic turn-on at lower masses (error function)
followed by a rapidly falling tail (exponential). The
pre-fit normalization is set to the LO \textsc{MadGraph}+{\PYTHIA}6 cross section
times an empirical factor of 1.3. This factor provides an initial
estimate of the difference between data and simulation in the
topologies, effectively accounting for the expected increase in the
inclusive cross section from NNLO corrections, and given a loose
$\pm$50\% constraint. The shape parameters of the function are allowed
to vary in the fit to the data without constraint.

The top quark background is a combination of \ttbar\ and single top quark
production processes.  The top quark model is parametrized by a sum of an error
function times an exponential function and a double Gaussian function, corresponding to
both merged and unmerged jets from hadronic \PW\ decays.  The top control sample is used to
correct the \PW\ resonance shape parameters, to estimate the expected
yield and yield uncertainties by extrapolating to the
signal region, and to adjust the top normalization uncertainty.  All top
shape parameters are fixed in the fit to the data, and the
normalization is constrained to a Gaussian with a width of 8 (10)\% for
muons (electrons). These come from a combination of theory uncertainty
and uncertainties associated with use of the top control sample.

Figure~\ref{fig:Fig1} shows the results of one of the fits to the data.
The left plots show the observed $m_{J}$ distributions, together
with the fitted contributions of the three largest SM
processes. The central plots show the same distribution after
subtracting all SM contributions from data except for diboson
events. The right plots show the pull distribution, i.e., the
normalized residual defined as
$(\text{data} - \text{fit})/(\text{fit uncertainty})$,
where the fit uncertainty is computed at each data point by
propagating the uncertainty in the normalization coefficients.

The
individual process yields, as determined by the fit, are reported in
Table~\ref{tab:yields}.  The acceptance times efficiency
($\mathcal{A}\varepsilon$) is determined from the diboson MC.  The electron
channel has a smaller $\mathcal{A}\varepsilon$ because of its higher
kinematic threshold.
The top quark results reflect the inability of the fit to further constrain
this background. The \PW+jets yields are about 20\% higher than the prefit
value of 1.3 times the LO prediction, which exhibits our limited knowledge
of this boosted regime. For the diboson process, 1.35 (2.23) times the expected
event count is observed in the muon (electron) channel. This excess is
statistically consistent with the SM NLO prediction~\cite{Alwall:2014hca}.
Overall, the approach produces a high quality model of the data
(Fig.~\ref{fig:Fig1}(left)), with pull distributions
consistent with zero (Fig.~\ref{fig:Fig1}(right)), that
allows us to extract the diboson contribution to the V$_\text{had}$ resonance
(Fig.~\ref{fig:Fig1}(center)).

\begin{table}
  \topcaption{Observed event yields and associated
  ratios (in parentheses) with respect to the pre-fit values
  extracted in the signal region
  ($70<m_J<100\GeV)$.
  The term $\mathcal{A}\varepsilon$ (acceptance $\times$ efficiency)
  includes \PW\ and \cPZ\ branching
  fractions~\cite{pdg}.}  \label{tab:yields} \centering {
  \begin{tabular}{lll}
\hline
      Quantity & $\Pgm$ channel & $\Pe$ channel \\
\hline
      Data      & 1977                   & 1666  \\
      \PW+jets  & 1318 ($1.22 \pm 0.06$) & 1023 ($1.17 \pm 0.07$) \\
      Top quark &  450 ($1.00 \pm 0.08$) &  364 ($1.00 \pm 0.10$) \\
\hline
      {\PW}V    &  204 ($1.35 \pm 0.77$) &  285 ($2.23 \pm 0.84$)  \\
$\mathcal{A}\varepsilon$ & $9.7\times 10^{-5}$ & $8.3\times 10^{-5}$  \\
\hline
\end{tabular}
}
\end{table}

\begin{figure}[htbp]
 {\centering
    \includegraphics[width=0.45\textwidth]{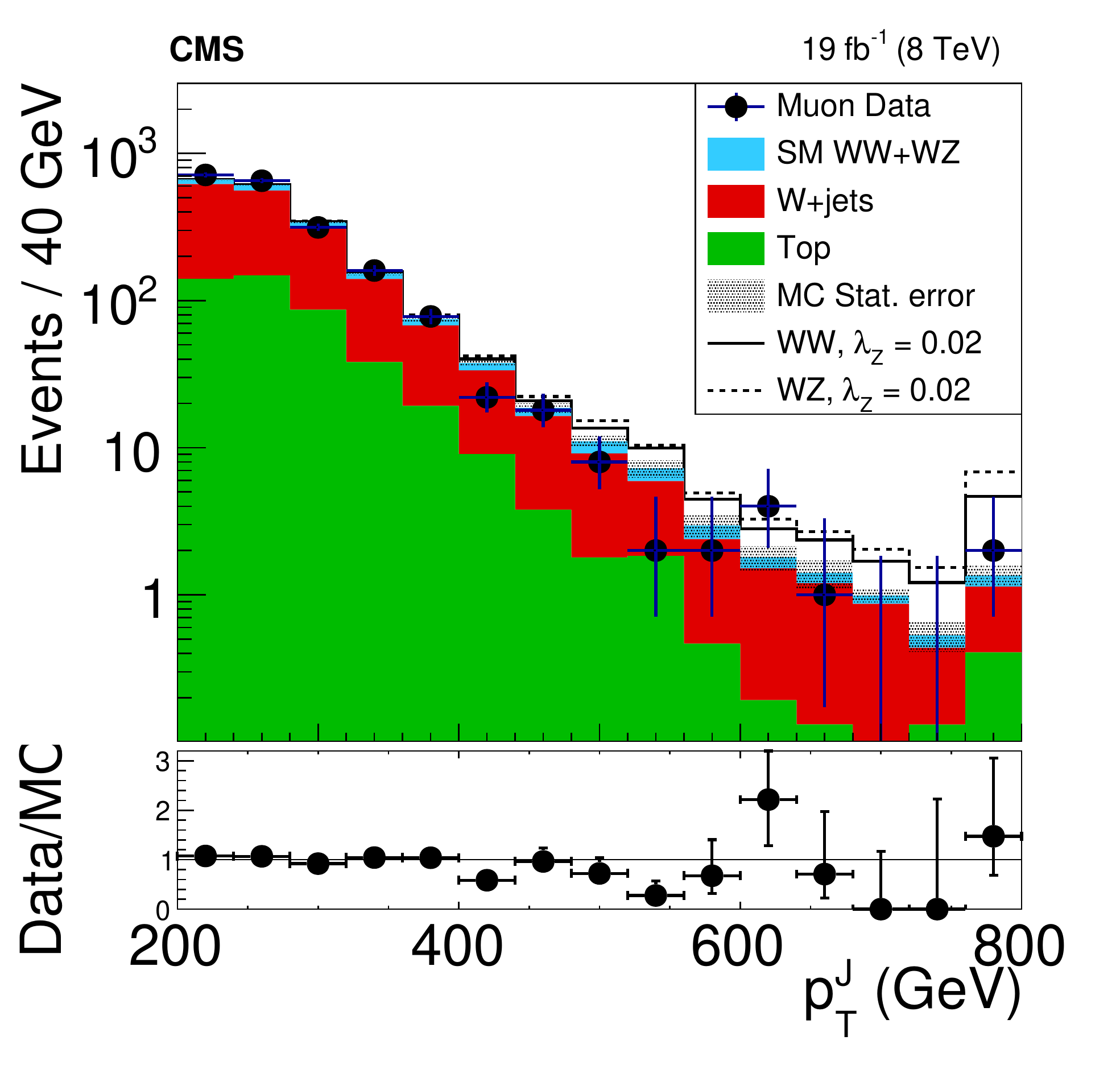}
    \includegraphics[width=0.45\textwidth]{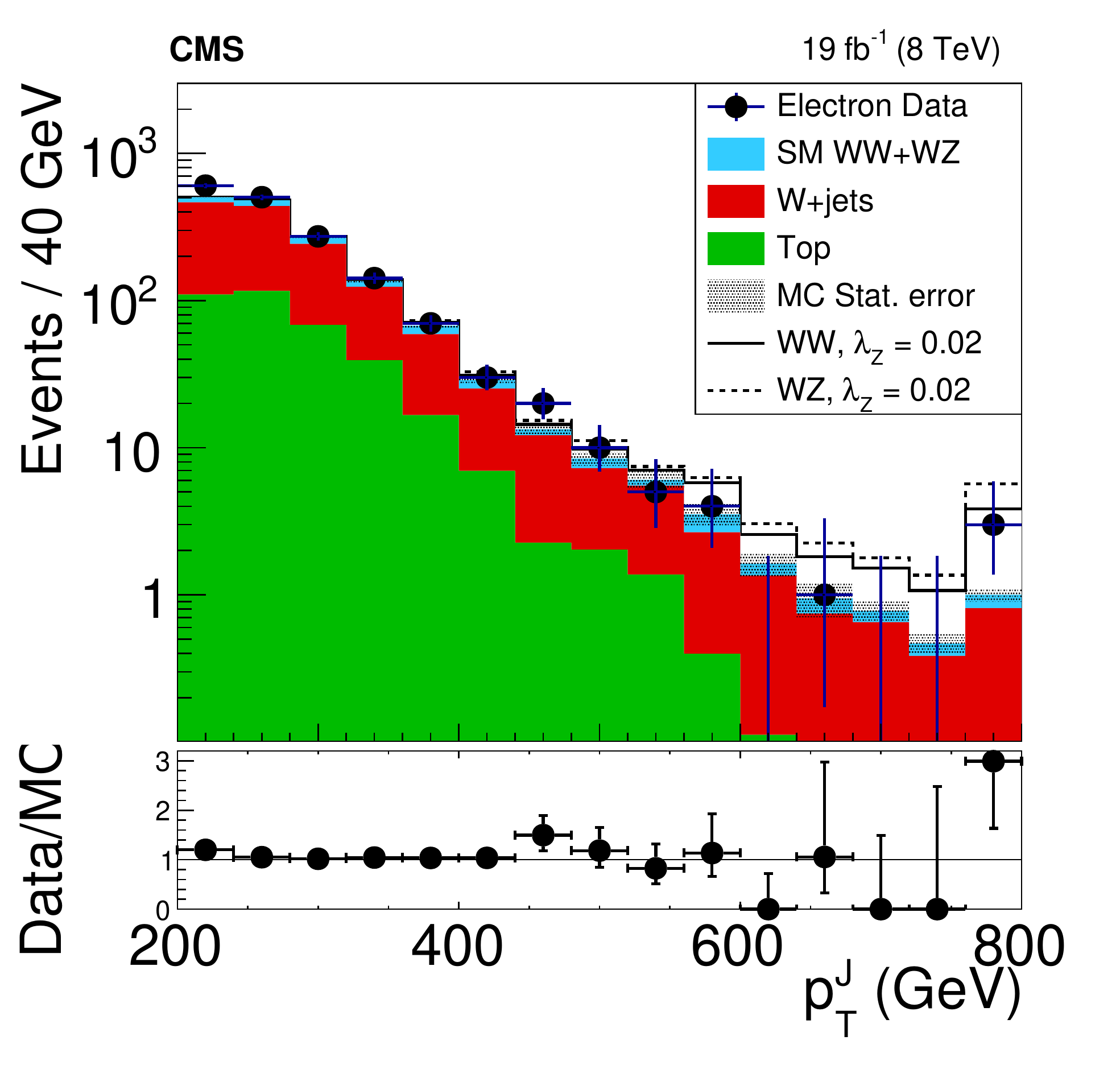}
    \caption{V$_\text{had}$ \pt distributions for the
    muon (\cmsLeft) and electron (\cmsRight) channels after full selection and with
    the requirement $70<m_{J}<100\GeV$.  The MC errors are
    purely statistical. Examples of the
    effects of aTGCs are shown by the solid and dotted lines.
    Below we show the data/MC ratio. The last bin
    includes the overflow. } \label{fig:Fig2} }
\end{figure}

\subsection{Fit validation}

We validate the fit procedure by performing
pseudo-experiments. For
each experiment, we generate the $m_{J}$ pseudo-data for the SM
processes using the fitted pdf, taking into
account the correlations between the yields, and then perform a fit to
each pseudo-data $m_{J}$ distribution as if it were the real
data. Likewise, we ensure that the parametrization used is
sufficiently general by generating pseudo-data with more general
functional forms and fitting them with the default configuration.
The results indicate that biases in all background yields and
yield uncertainties are small.

\subsection{Signal modeling}
\label{sec:sigmod}

The dependence of the $\ptJ$ distribution on specific aTGCs is
modeled by reweighting the simulations of SM \PW\PW\ and \PW\cPZ\
by the ratio of squared matrix elements
with and without the anomalous coupling, i.e., $\abs{{\mathcal
M}}^{2}/\abs{{\mathcal M}}_\mathrm{SM}^{2}$, where $\abs{{\mathcal M}}^{2}$ is the
squared matrix element in the presence of anomalous couplings and
$\abs{{\mathcal M}}_\mathrm{SM}^{2}$ is the squared matrix element in the SM,
calculated with \MCFM version 6.0~\cite{MCFM}. These ratios are calculated,
parametrized with polynomials, and the polynomials encapsulated into a
unified signal model in two-dimensional (2D) space for three pairwise
combinations of the effective Lagrangian parameters being studied.

\subsection{Preparing \texorpdfstring{$\ptJ$}{merged jet pt} distributions}
\label{sec:prepdist}

Distributions of $\ptJ$ in the form of histograms binned over the
range 200--800\GeV (Fig.~\ref{fig:Fig2}) are used to compute limits. All selections
have been applied, including the signal window, $70 <
m_{J}<100\GeV$. The \PW+jets and top quark background normalizations
are fixed according to the results from the $m_{J}$ fits. The SM
diboson components, however, are normalized to the NLO
predictions, since a) we are searching for enhancements to the diboson
production relative to those predictions, and b) given the excess of SM diboson events
obtained from the fits in both channels, normalizing to theory predictions yields
substantially more conservative, less sensitive expected limits. We treat
the two lepton categories as separate channels in the limit setting process.

Since the \PW+jets shape is only calculated to LO, and we are
exploring a new region of phase space, we adjust the shape and
normalization from MC by comparing it to a distribution derived using
an alternative method.  This method involves extrapolating
the \PW+jets $\ptJ$ distribution from a $m_{J}$ data sideband
to the signal region by means of a transfer
function. The transfer function is a ratio of curves fitted to
the \PW+jets $\ptJ$ distributions in the signal and sideband
regions of \PW+jets
simulation~\cite{Khachatryan:2014gha,Khachatryan:2015cwa}. The
comparison shows that the ratio of the \PW+jets backgrounds
derived using the two methods is statistically consistent with unity.

\section{Systematic uncertainties}

The main source of systematic uncertainty is the normalization
uncertainty in the \PW+jets background estimate. From the alternative
method described in Sec.~\ref{sec:prepdist}, we extract a 20\% uncertainty in the total background
normalization by taking the precision of the ratio of the \PW+jets
background distribution derived from the two methods and summing
over the high \pt region (400--800\GeV), where the signal is expected.

The theoretical uncertainties in the signal normalization are
associated with the renormalization and factorization scales, and with the choice of
PDF, for $\pt^\PW>1\TeV$.
For PDF uncertainties we compare a\MCATNLO samples employing
41 alternative sets of CTEQ6M PDFs following the prescription in
Ref.~\cite{Pumplin:2002vw}.  Factorization and renormalization scale
uncertainties are estimated by simultaneously varying them up or down
by a factor of 2. Both scale and PDF uncertainties are estimated to be
approximately 18--26\%.

The uncertainty in the signal shape coming from the effects of reconstruction
is estimated by comparing the aTGC/SM ratios at the generator level and the
aTGC/SM ratios at the reconstruction level after all major selections
are applied for both samples.  The ratio is consistent with unity, and therefore
only the statistical error on the ratio is propagated as an
uncertainty in the modeling of different aTGC signal grid points.

The uncertainty in the luminosity measurement is 2.6\%~\cite{lumiPAS}.
Additional sources of uncertainty from limited MC
sample size, jet energy scale and resolution, \MET resolution, trigger
efficiency, lepton reconstruction and selection efficiency, additional
jet veto, pileup, and b-tag efficiency are negligible in comparison to
the primary sources. These uncertainties are
treated as nuisance parameters in the model and profiled according
to Ref.~\cite{LHC-HCG}, Appendix~A. Luminosity and theory uncertainties
are treated as 100\% correlated between the two channels.

\section{Coupling limits and summary}

\begin{figure*}[tbh]
  {\centering
  \includegraphics[width=\tripleColFig]{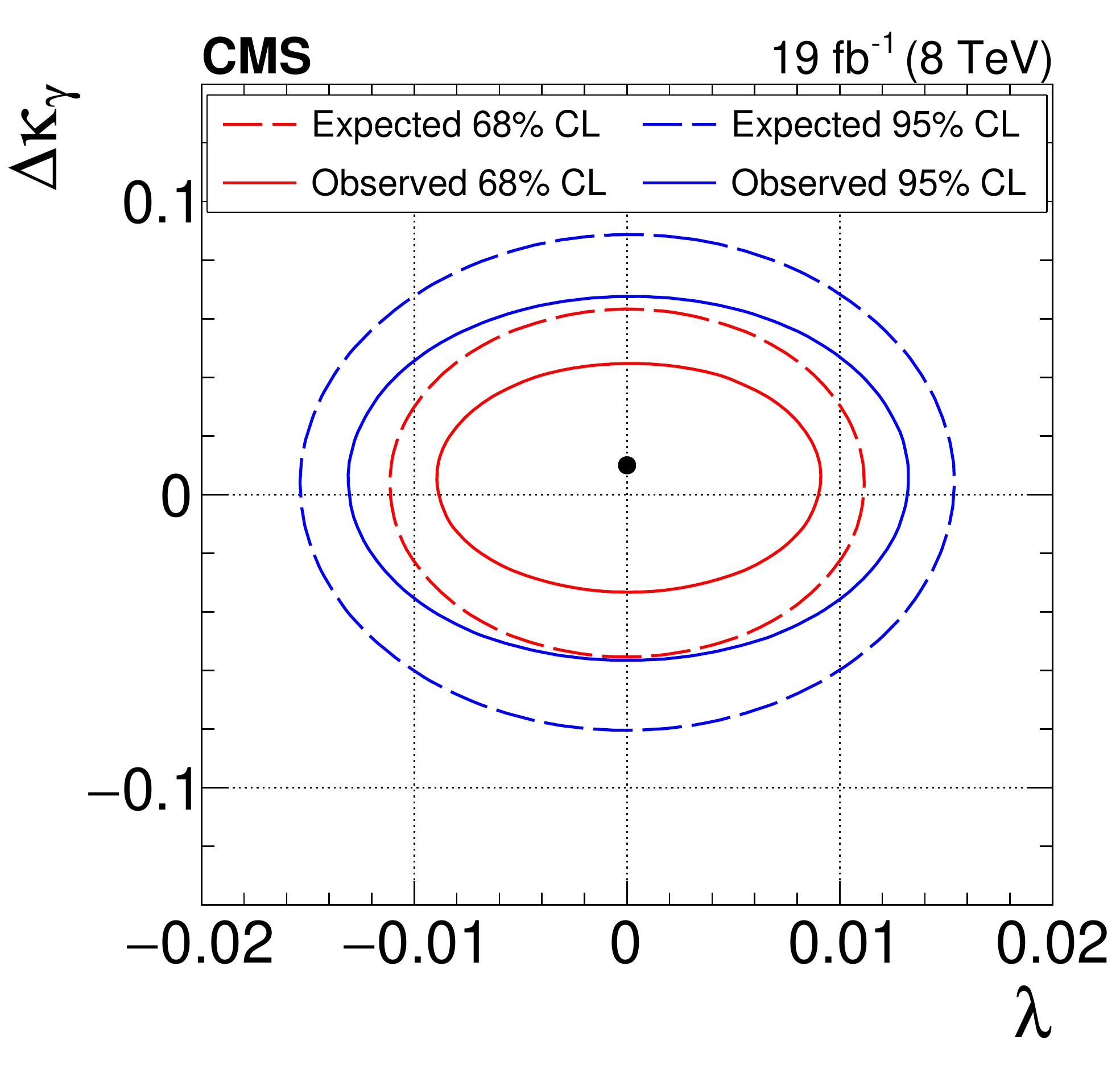}
  \includegraphics[width=\tripleColFig]{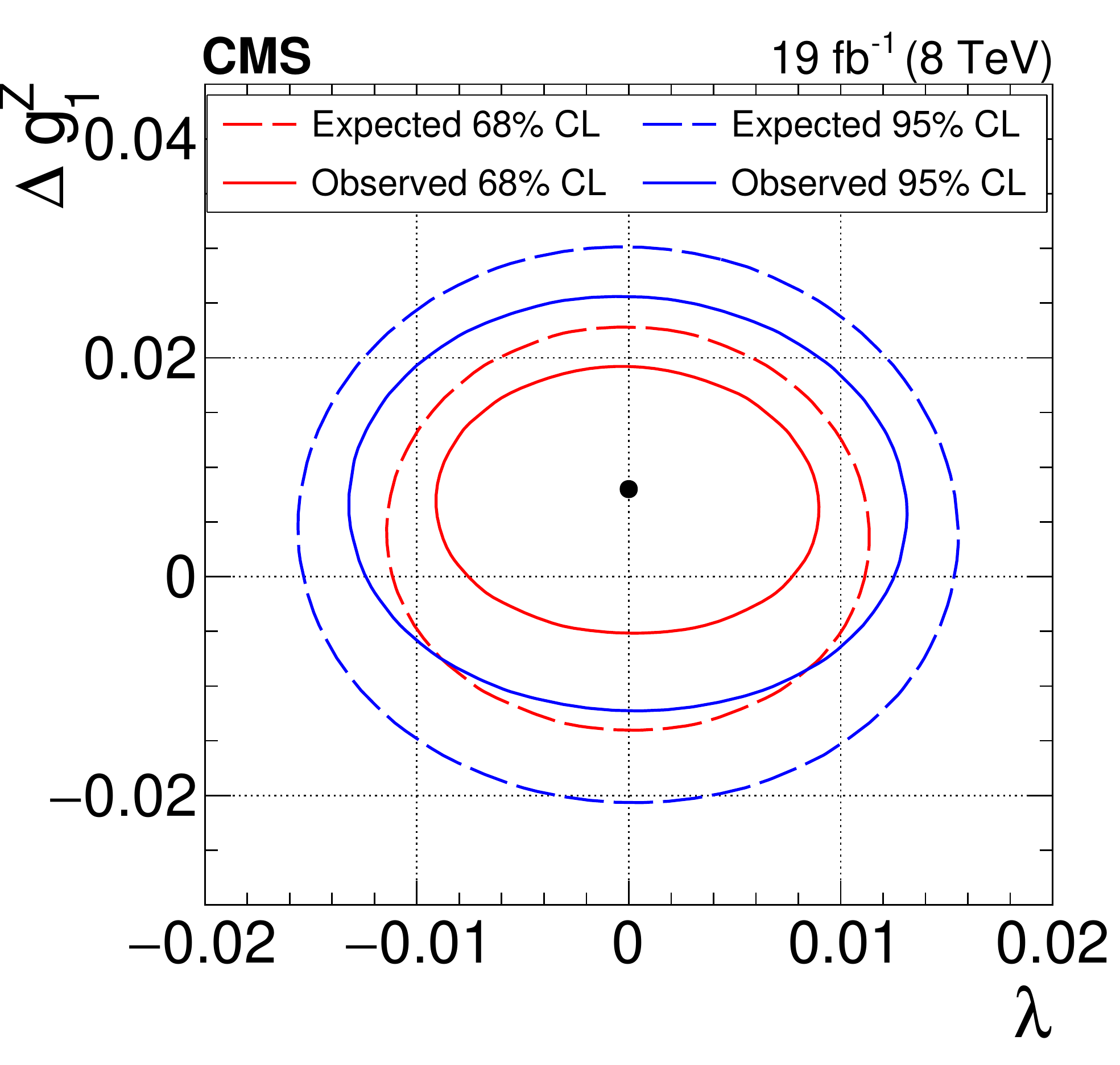}
  \includegraphics[width=\tripleColFig]{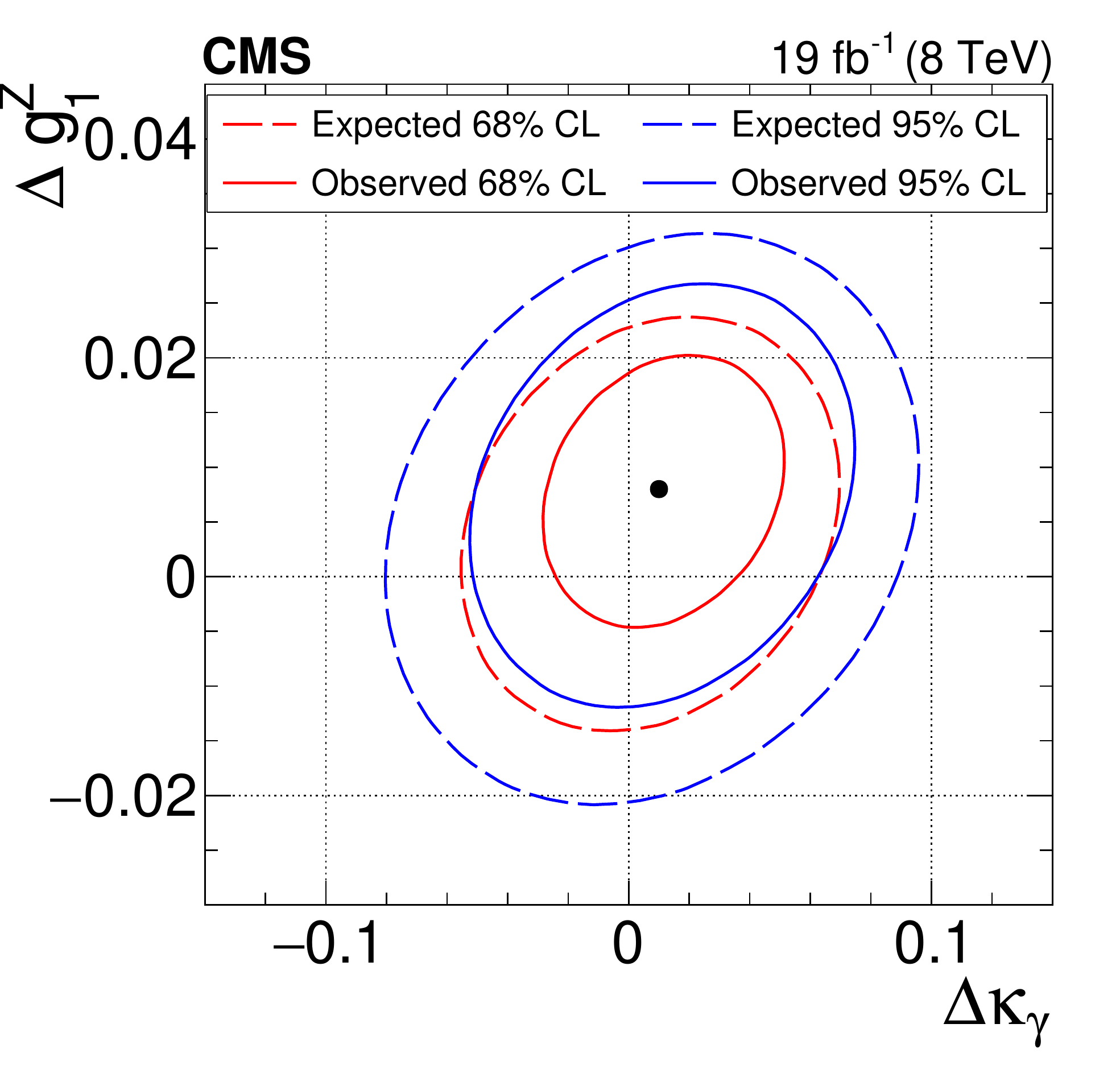} \\
  \includegraphics[width=\tripleColFig]{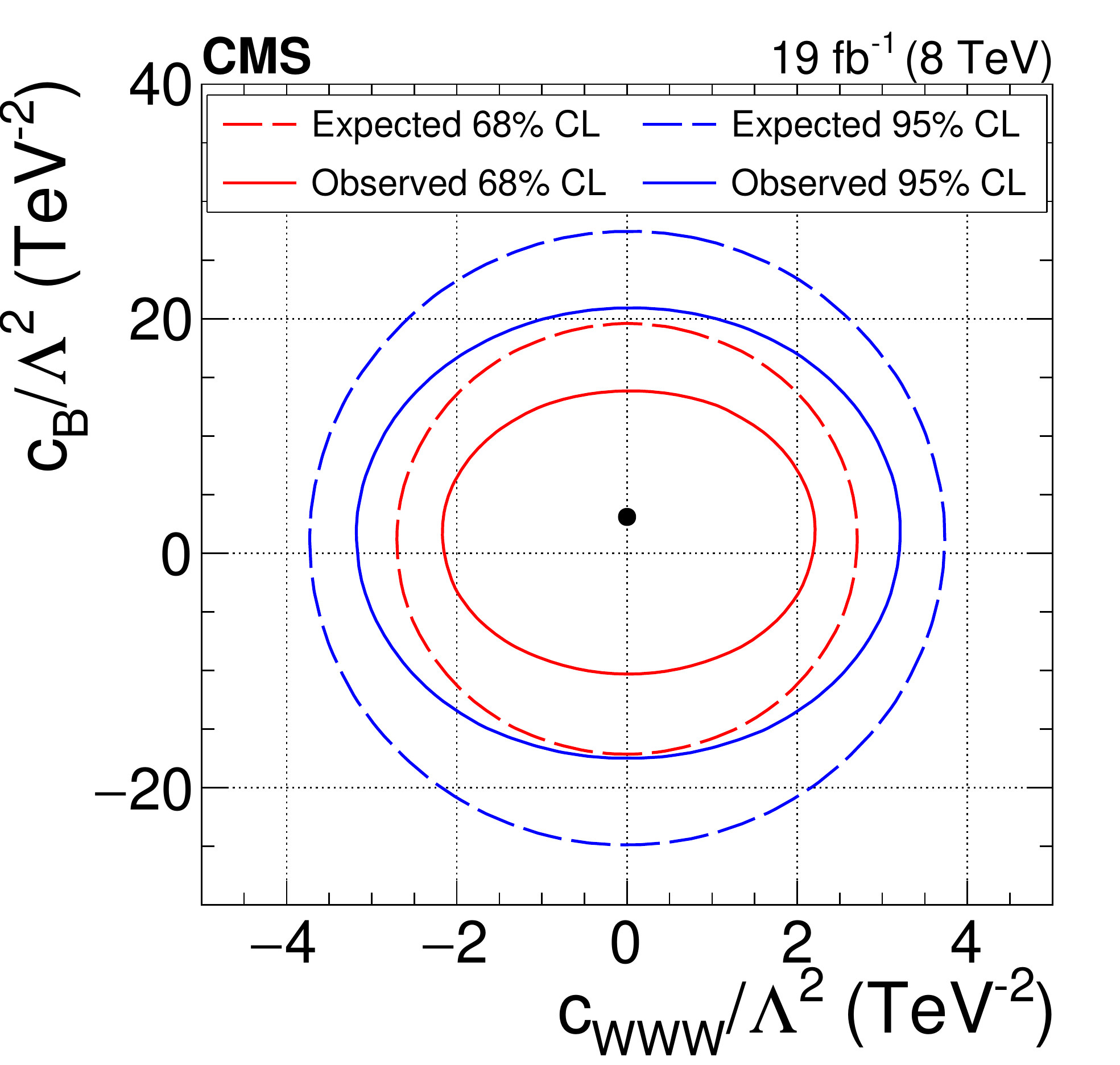}
  \includegraphics[width=\tripleColFig]{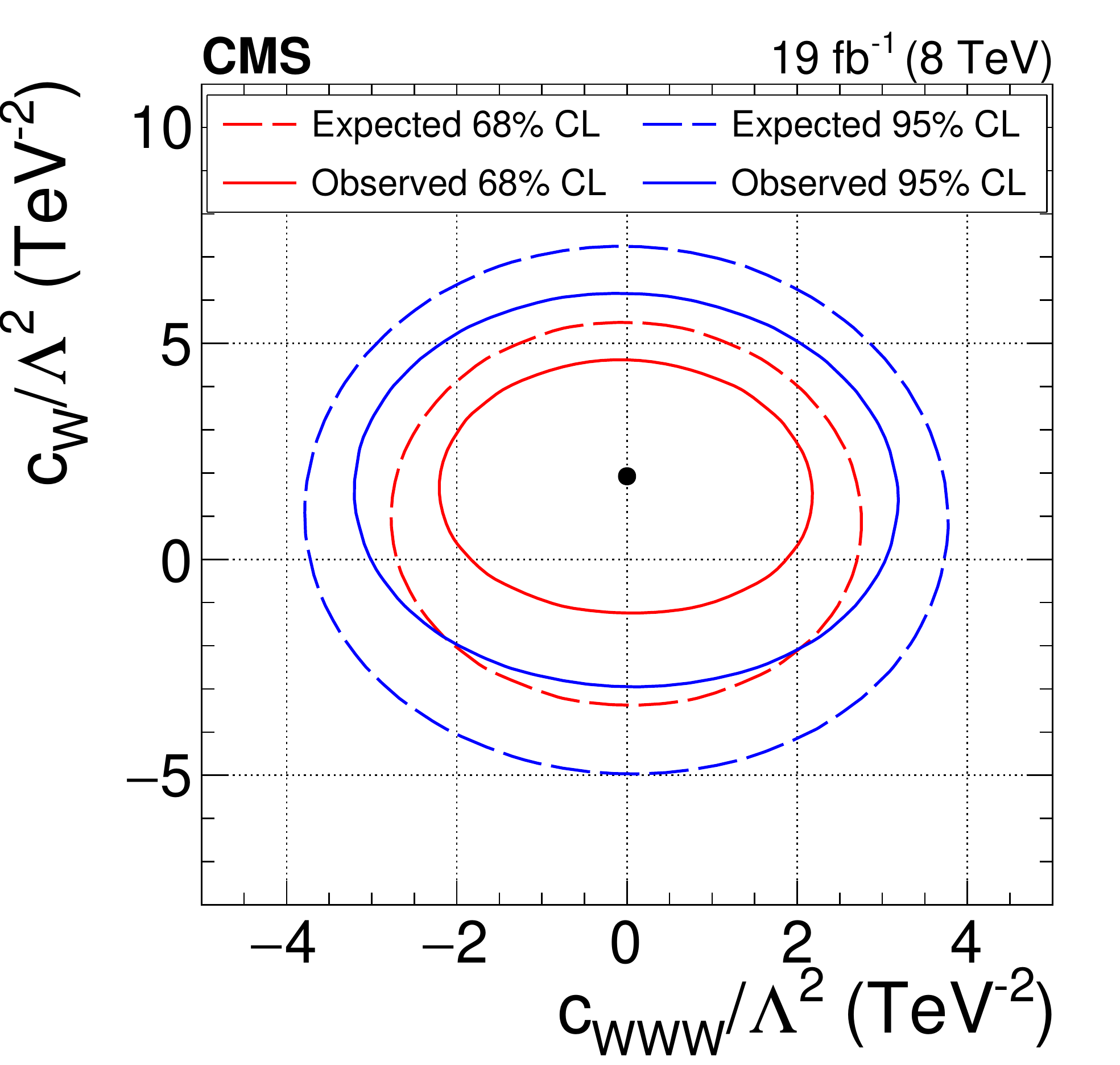}
  \includegraphics[width=\tripleColFig]{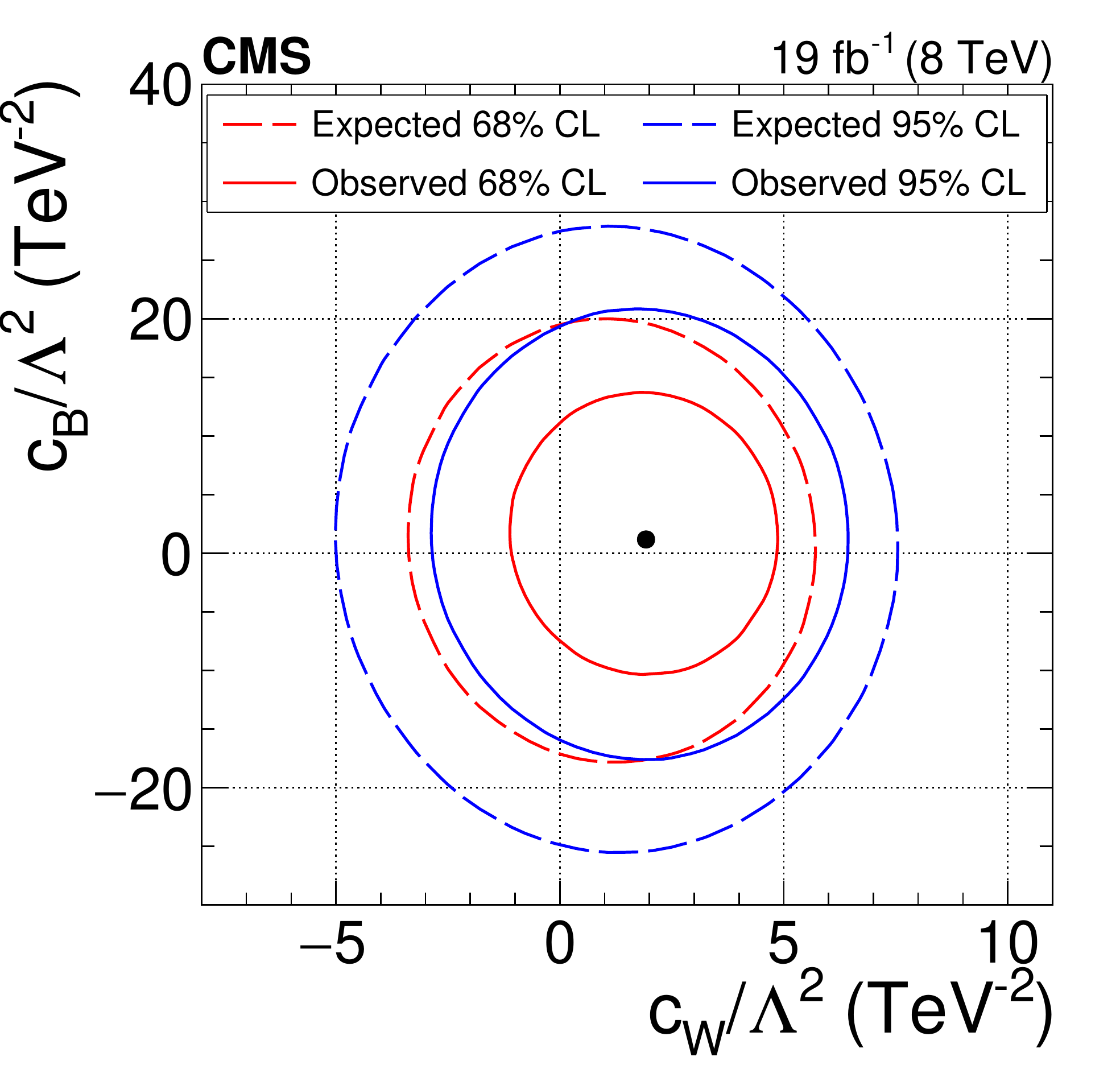}
    \caption{\label{fig:atgclimits}
    The 68 and 95\% CL observed and expected
    exclusion contours in $\Delta$NLL are depicted for three pairwise combinations of the
    aTGC parameters in the LEP parametrization (top)
    and in the EFT formulation (bottom). The black dot represents the best fit point.
    The origin represents the SM prediction. The asymmetry of expected limits around the
    SM is allowed by the theoretical parametrization. }
  }
\end{figure*}

Two-dimensional likelihood fits are performed in the three
planes described in Sec.~\ref{sec:sigmod}. Each time the third parameter is profiled.
The electron and muon channels are fitted simultaneously in the limit setting procedure. No
evidence for anomalous couplings is found, and we calculate the 68 and 95\%
confidence level (CL) exclusion contours, using the differences of the negative log
likelihood ($\Delta$NLL) relative to the best fit point.
No form factors are used.
The limits are subsequently translated~\cite{Degrande:2012wf}
into equivalent limits on the
parameters within the EFT approach, namely
$c_{\PW\PW\PW}/\Lambda^2$, $c_{\PW}/\Lambda^2$, and $c_\mathrm{B}/\Lambda^2$, shown
in Fig.~\ref{fig:atgclimits}.  We also set 1D 95\% CL limits on all
six parameters, with the second parameter profiled
and the third parameter fixed to zero.
These are shown in Tables~\ref{tab:limits1d} and~\ref{tab:eftlimits1d}.
The latter also shows other recent 8\TeV results for comparison.

In summary, our
limits are consistent with the SM prediction and improve upon the
sensitivity of the fully leptonic 8\TeV results~\cite{Khachatryan:2015sga,Aad:2016wpd}
and the combined LEP experiments~\cite{pdg,Schael:2013ita}.

\begin{table}
\begin{center}
\topcaption{Summary of expected and observed one-dimensional limits in the LEP parametrization.
Each number pair represents the observed 95\% confidence interval for
that parameter. \label{tab:limits1d}}
\begin{tabular}{ l  c  c  }
\hline
Parameter & Expected Limits & Observed Limits \\
\hline
$\lambda_{\cPZ}$     & $[ -0.014,  0.013 ]$ & $[ -0.011,  0.011 ]$ \\
$\Delta{\kappa_\gamma}$  & $[ -0.068,  0.082 ]$ & $[ -0.044,  0.063 ]$ \\
$\Delta{g_1^{\cPZ}}$ & $[ -0.018,  0.028 ]$ & $[ -0.0087, 0.024 ]$ \\
\hline
\end{tabular}
\end{center}
\end{table}

\begin{table}
\begin{center}
\topcaption{Summary of one-dimensional limits in the EFT formulation for this analysis (*)
compared to previous results. \label{tab:eftlimits1d}}
\begin{tabular}{ c  c  c  c }
\hline
  & $c_{\PW\PW\PW}/\Lambda^2$ & $c_\mathrm{B}/\Lambda^2$ & $c_{\PW}/\Lambda^2$ \\
  &  ($\TeV^{-2}$)             & ($\TeV^{-2}$)            & ($\TeV^{-2}$)            \\
\hline
*                          & $[ -2.7,  2.7  ]$ & $[ -14,   17   ]$ & $[  -2.0,   5.7 ]$ \\
\cite{Khachatryan:2015sga} & $[ -5.7,  5.9  ]$ & $[ -29.2, 23.9 ]$ & $[ -11.4,   5.4 ]$ \\
\cite{Aad:2016wpd}         & $[ -4.61, 4.60 ]$ & $[ -20.9, 26.3 ]$ & $[  -5.87, 10.54 ]$  \\
\cite{Khachatryan:2016poo} & $[ -4.6,  4.2  ]$ & $[ -260, 210   ]$ & $[  -4.2,   8.0 ]$   \\
\cite{Aad:2016ett}         & $[ -3.9,  4.0  ]$ & $[ -320, 210   ]$ & $[  -4.3,   6.8 ]$   \\
\hline

\end{tabular}
\end{center}
\end{table}

\begin{acknowledgments}
We congratulate our colleagues in the CERN accelerator departments for the excellent performance of the LHC and thank the technical and administrative staffs at CERN and at other CMS institutes for their contributions to the success of the CMS effort. In addition, we gratefully acknowledge the computing centers and personnel of the Worldwide LHC Computing Grid for delivering so effectively the computing infrastructure essential to our analyses. Finally, we acknowledge the enduring support for the construction and operation of the LHC and the CMS detector provided by the following funding agencies: BMWFW and FWF (Austria); FNRS and FWO (Belgium); CNPq, CAPES, FAPERJ, and FAPESP (Brazil); MES (Bulgaria); CERN; CAS, MoST, and NSFC (China); COLCIENCIAS (Colombia); MSES and CSF (Croatia); RPF (Cyprus); SENESCYT (Ecuador); MoER, ERC IUT, and ERDF (Estonia); Academy of Finland, MEC, and HIP (Finland); CEA and CNRS/IN2P3 (France); BMBF, DFG, and HGF (Germany); GSRT (Greece); OTKA and NIH (Hungary); DAE and DST (India); IPM (Iran); SFI (Ireland); INFN (Italy); MSIP and NRF (Republic of Korea); LAS (Lithuania); MOE and UM (Malaysia); BUAP, CINVESTAV, CONACYT, LNS, SEP, and UASLP-FAI (Mexico); MBIE (New Zealand); PAEC (Pakistan); MSHE and NSC (Poland); FCT (Portugal); JINR (Dubna); MON, RosAtom, RAS, RFBR and RAEP (Russia); MESTD (Serbia); SEIDI, CPAN, PCTI and FEDER (Spain); Swiss Funding Agencies (Switzerland); MST (Taipei); ThEPCenter, IPST, STAR, and NSTDA (Thailand); TUBITAK and TAEK (Turkey); NASU and SFFR (Ukraine); STFC (United Kingdom); DOE and NSF (USA).

\hyphenation{Rachada-pisek} Individuals have received support from the Marie-Curie program and the European Research Council and EPLANET (European Union); the Leventis Foundation; the A. P. Sloan Foundation; the Alexander von Humboldt Foundation; the Belgian Federal Science Policy Office; the Fonds pour la Formation \`a la Recherche dans l'Industrie et dans l'Agriculture (FRIA-Belgium); the Agentschap voor Innovatie door Wetenschap en Technologie (IWT-Belgium); the Ministry of Education, Youth and Sports (MEYS) of the Czech Republic; the Council of Science and Industrial Research, India; the HOMING PLUS program of the Foundation for Polish Science, cofinanced from European Union, Regional Development Fund, the Mobility Plus program of the Ministry of Science and Higher Education, the National Science Center (Poland), contracts Harmonia 2014/14/M/ST2/00428, Opus 2014/13/B/ST2/02543, 2014/15/B/ST2/03998, and 2015/19/B/ST2/02861, Sonata-bis 2012/07/E/ST2/01406; the National Priorities Research Program by Qatar National Research Fund; the Programa Clar\'in-COFUND del Principado de Asturias; the Thalis and Aristeia programs cofinanced by EU-ESF and the Greek NSRF; the Rachadapisek Sompot Fund for Postdoctoral Fellowship, Chulalongkorn University and the Chulalongkorn Academic into Its 2nd Century Project Advancement Project (Thailand); and the Welch Foundation, contract C-1845.
\end{acknowledgments}
\bibliography{auto_generated}

\cleardoublepage \appendix\section{The CMS Collaboration \label{app:collab}}\begin{sloppypar}\hyphenpenalty=5000\widowpenalty=500\clubpenalty=5000\textbf{Yerevan Physics Institute,  Yerevan,  Armenia}\\*[0pt]
A.M.~Sirunyan, A.~Tumasyan
\vskip\cmsinstskip
\textbf{Institut f\"{u}r Hochenergiephysik,  Wien,  Austria}\\*[0pt]
W.~Adam, E.~Asilar, T.~Bergauer, J.~Brandstetter, E.~Brondolin, M.~Dragicevic, J.~Er\"{o}, M.~Flechl, M.~Friedl, R.~Fr\"{u}hwirth\cmsAuthorMark{1}, V.M.~Ghete, C.~Hartl, N.~H\"{o}rmann, J.~Hrubec, M.~Jeitler\cmsAuthorMark{1}, A.~K\"{o}nig, I.~Kr\"{a}tschmer, D.~Liko, T.~Matsushita, I.~Mikulec, D.~Rabady, N.~Rad, B.~Rahbaran, H.~Rohringer, J.~Schieck\cmsAuthorMark{1}, J.~Strauss, W.~Waltenberger, C.-E.~Wulz\cmsAuthorMark{1}
% ...
\vskip\cmsinstskip
% ...
1:~~Also at Vienna University of Technology, Vienna, Austria\\
% ...

\end{sloppypar}
\end{document}